\documentclass[12pt]{article}

\usepackage{graphicx}
\usepackage{epsf}

\usepackage{amsmath, amssymb}

\begin{document}
  
\begin{titlepage}

\begin{center}

\hfill TU-854\\
\hfill Octorber, 2009

\vspace{1.5cm}
{\Large\bf 
Measurement of the Superparticle Mass Spectrum
\\
in the Long-Lived Stau Scenario at the LHC
}

\vspace{1cm}
{\large Takumi Ito, Ryuichiro Kitano and Takeo Moroi}

\vspace{1cm}

{\it {Department of Physics, Tohoku University,
    Sendai 980-8578, Japan}}

\vspace{1cm}
\abstract{ 
  
  In supersymmetric scenarios with a long-lived stau, the LHC
  experiments provide us with a great environment for precise mass
  measurements of superparticles.  We study a case in which the mass
  differences between the lightest stau and other sleptons are $\sim
  10\ {\rm GeV}$ or larger, so that the decay products of heavier
  sleptons are hard enough to be detected. We demonstrate that the
  masses of neutralinos, sleptons, and squarks can be measured with a
  good accuracy.

 }

\end{center}
\end{titlepage}

\renewcommand{\theequation}{\thesection.\arabic{equation}}
\renewcommand{\thepage}{\arabic{page}}
\setcounter{page}{1}
\renewcommand{\thefootnote}{\#\arabic{footnote}}
\setcounter{footnote}{0}

\section{Introduction}
\label{sec:intro}
\setcounter{equation}{0}

Signatures of supersymmetry (SUSY) at the LHC experiments crucially
depend on what the lightest supersymmetric particle (LSP) is. Many of
studies have assumed a neutralino as the LSP motivated by a possible
explanation of dark matter of the universe. In this case, final states
of SUSY events will be accompanied by missing momentum carried away by
two neutralinos. The mass determinations of SUSY particles in such cases
are not a straightforward task. One needs to combine various
measurements to extract the masses out of observables.

Situation drastically changes when we assume that the scalar tau lepton
(stau) is lighter than the lightest neutralino and thus
long-lived. 
First, one can discover the stau by looking for anomalous tracks at the
inner tracker and the muon detector~\cite{Drees:1990yw, Feng:1997zr,
Martin:1998vb}.
Their momenta and velocities can be measured by analyzing the tracks,
with which one can measure the stau mass with a good
accuracy~\cite{Nisati:1997gb, atlasnote, Ambrosanio:2000ik,
  ellis_atlnote}. The momentum information enables us to perform
precise determination of the properties of SUSY particles such as
masses of other superparticles~\cite{Hinchliffe:1998ys, Ellis:2006vu,
  Ibe:2007km, Feng:2009yq}, the spin of the
stau~\cite{Rajaraman:2007ae}, the lifetime of the
stau~\cite{Buchmuller:2004rq, Hamaguchi:2004df, Feng:2004yi,
  Ishiwata:2008tp} and P/CP/T violations in the SUSY
interactions~\cite{Kitano:2008sa}.

The long-lived stau is not just motivated as a golden scenario for
collider experiments. There are many underlying models and parameter
spaces in them to realize the scenario. Examples include the familiar
ones such as supergravity and gauge mediation models.  The lifetime of
stau depends on details of the microscopic models; the stau can decay
into a gravitino and a tau lepton if kinematically allowed or into
standard model particles if $R$-parity is violated. There is a
cosmological constraint on the stau lifetime from Big-Bang
nucleosynthesis, but it can easily be evaded unless the lifetime is
extremely long such as greater than $O(1000\ {\rm sec})$
\cite{Kawasaki:2008qe}.

Once the stau tracks are discovered and the measurement of the stau
mass is done at the LHC, the next important quantities to measure will
be other superparticle masses.  In particular, measurements of
selectron and smuon masses tell us about how special the third
generation is in the microscopic interactions between the SUSY
breaking and the standard model sectors.  Even if the strength of the
Yukawa interaction is the only difference between the stau and the
first two generations, the mass splitting contains the information on
the size of the Yukawa interaction (i.e., $\tan \beta$) and also the
size of the quantum corrections to the masses (i.e., the messenger
scale).  An even more interesting quantity will be the mass splitting
between the selectron and the smuon. Since the Yukawa coupling
constants are small, a large splitting of more than $O(100\ {\rm
  MeV})$ level requires an explanation from a microscopic theory.

The mass measurements of superparticles in the long-lived stau
scenario was first studied in \cite{Hinchliffe:1998ys}. They used a
parameter point in the gauge-mediation model where the selectron, the
smuon, and the stau are all long-lived due to a small mass splittings
among three sleptons ($m_{\tilde e} \sim m_{\tilde \mu} \sim m_{\tilde
  \tau}$). Recently, the case with a long-lived selectron with a
nearly degenerate smuon ($m_{\tilde e} \sim m_{\tilde \mu} < m_{\tilde
  \tau}$) was studied in \cite{Feng:2009yq}, there it was proposed to
measure the mass splitting between the smuon and the selectron by
looking at the peak locations of the $e\tilde e$ and $\mu\tilde e$
invariant masses.

In this paper, we propose a method to measure the masses of
superparticles in the case where $m_{\tilde \tau} < m_{\tilde e} \sim
m_{\tilde \mu} < m_{\chi^0}$ in the long-lived stau scenario at the
LHC.  When the mass differences between the stau and other sleptons
are larger than $\sim 10\ {\rm GeV}$, the decay products are hard
enough to be detected, contrasted to the previous studies. This level
of mass splitting is a natural expectation in many models especially
for large values of the $\tan\beta$ parameter (which is somewhat
motivated by the Higgs boson mass bound).

At the LHC, staus are mainly produced at the last step of the
cascade-decay chains such as the neutralino decay $\chi^0 \to \tau
\tilde \tau$ and the chargino decay $\chi^\pm \to \tilde \tau \nu$.
It has been demonstrated in various models that the neutralino
mass(es) can be measured by looking for an endpoint of the
$j_\tau$-$\tilde \tau$ invariant mass
distribution~\cite{Hinchliffe:1998ys, Ellis:2006vu, Ibe:2007km} (where
$j_\tau$ is $\tau$-tagged jet).  We first follow this analysis and
point out usefulness of the charge information of the $\tau$-jets in
this study.  We next show that the measured neutralino mass in turn
can be used to determine the smuon and selectron masses by
reconstructing the decay chains: $\chi^0 \to \mu \tilde \mu \to \mu
(\tilde \tau \tau \mu)$ and $\chi^0 \to e \tilde e \to e (\tilde \tau
\tau e)$ followed by a $\tau$ decay. We use the hadronic $\tau$ decay
for the analysis.  Although there are invisible neutrinos from $\tau$
decays, four-momentum of $\tau$ can be reconstructed in the
event-by-event basis by using the knowledge of the neutralino mass
under an assumption that the neutrino is emitted along the direction
of the $\tau$-jet (which is valid when $\tau$ is highly boosted). The
smuon and selectron masses can then be measured in each event up to a
combinatorics. Therefore, we can see sharp peaks at the smuon and
selectron masses in the distributions of the $\tilde \tau \tau \mu$
and $\tilde \tau \tau e$ invariant masses, respectively.  We
demonstrate that the masses and the mass difference can be measured
with accuracies of $O(100~{\rm MeV})$. This level of accurate mass
measurements will provide us with a very important information on the
underlying microscopic theory.
We also proceed to reconstruct squark masses by using the $\tilde q
\to q \chi_1^0 \to q \tau\tilde{\tau}$ decay.  We can see a sharp peak
in the $q$-$\tau$-$\tilde{\tau}$ invariant mass distribution.

The organization of this paper is as follows.  In Section
\ref{sec:idea}, basic ideas to measure the neutralino, slepton and
squark masses are explained.  In order to demonstrate that our basic
ideas work, we show the results of Monte Carlo (MC) analysis in
Section~\ref{sec:mc}.
We will discuss implications to microscopic theories in
Section~\ref{sec:imp}.
Section~\ref{sec:conclusions} is devoted to conclusions.

\section{Basic Ideas}
\label{sec:idea}
\setcounter{equation}{0}

As mentioned in Introduction, in a class of SUSY models, the lightest
stau $\tilde{\tau}_1$ can be the lightest superparticle in the MSSM
sector (which we call the MSSM-LSP).  The lifetime of $\tilde{\tau}_1$
can be long enough to escape the detector before their decays, thereby
one can treat it as a stable particle in collider experiments.

If $\tilde{\tau}_1$ is long-lived, we expect a unique LHC phenomenology,
very different from the case of the neutralino LSP.  
In particular, we will see tracks of $\tilde{\tau}_1$ and thus the
momentum information on the MSSM-LSP will be available in contrast to
the missing momentum in the neutralino LSP case. The momentum
information enables us to easily reconstruct SUSY events.
It is also notable that, once the $\tilde{\tau}_1$ track is identified
in an event by looking at both the velocity and the momentum, the event
can be distinguished from the standard-model events, which in principle
eliminates standard-model backgrounds.
It has been studied that the selection of slow tracks effectively reduce
the muon background~\cite{Ambrosanio:2000ik,ellis_atlnote}.
In the following analysis, we assume that the $\tilde{\tau}_1$ track can
be distinguished from the muon track if $\beta_{\tilde{\tau}_1}\leq
0.9$~\cite{Ambrosanio:2000ik}, and we neglect the standard-model
background.

Collider phenomenology of the long-lived $\tilde{\tau}_1$ scenario will
be quite different depending upon the mass spectrum of the
superparticles.  In this paper, we consider the case with the following
mass relation:
\begin{eqnarray}
  m_{\tilde{\tau}_1} < m_{\tilde{e}_R, \tilde{\mu}_R} < m_{\chi^0_1},
  \label{MassRelation}
\end{eqnarray}
where $m_{\tilde{\tau}_1}$, $m_{\tilde{e}_R}$, $m_{\tilde{\mu}_R}$,
and $m_{\chi^0_1}$ are masses of $\tilde{\tau}_1$, lighter selectron
$\tilde{e}_R$, lighter smuon $\tilde{\mu}_R$, and the lightest
neutralino $\chi^0_1$, respectively.  In addition, we assume the
following for simplicity:
\begin{itemize}
\item All the colored SUSY particles are heavier than $\chi^0_1$.
\item The lighter sleptons are (almost) right-handed.
\item The heavier sleptons, which are almost left-handed, are heavier
  than $\chi^0_1$.
\end{itemize}
We pay particular attention to the case that the mass differences
$m_{\tilde{e}_R, \tilde{\mu}_R}-m_{\tilde{\tau}_1}$ and
$m_{\chi^0_1}-m_{\tilde{e}_R, \tilde{\mu}_R}$ are both sizable; if so,
the decay products of $\chi^0_1$ and sleptons are energetic enough to
be detected.  These conditions are realized in a wide class of SUSY
breaking models; one of the examples is the minimal gauge-mediated
SUSY breaking (GMSB) model~\cite{Dine:1994vc} with a large value of
$\tan\beta$.

With the mass spectrum mentioned above, some of the neutralinos (in
particular, $\chi^0_1$) decay as $\chi^0_i\rightarrow
\tau^\pm\tilde{\tau}_1^\mp$ and $\chi^0_i\rightarrow
l^\pm\tilde{l}_R^\mp$.  (Here and hereafter, $l$ stands for $e$ and
$\mu$.)  In the latter case, $\tilde{l}_R$ then decays into three-body
final state: $\tilde{l}_R\rightarrow l\tau^\pm\tilde{\tau}_1^\mp$.
Since the momentum of $\tilde{\tau}_1$ can be measured, we can
reconstruct $m_{\tilde{e}_R}$, $m_{\tilde{\mu}_R}$, and $m_{\chi^0_i}$
using these decay processes.

The basic procedures are as follows.  For the reconstruction of the
neutralino mass $m_{\chi^0_i}$, one can use the decay process
$\chi^0_i\rightarrow \tau^\pm\tilde{\tau}^\mp_1$, followed by the
hadronic decay of $\tau$.
Since $\tau$-jets have a distinguishable feature from the typical QCD
jets; a very narrow jet containing small number of charged track(s),
one can identify it with a high efficiency. (Here, we only use 1- and
3-prong decay of $\tau$.)
If we consider the distribution of an invariant mass:
$M_{j_\tau\tilde{\tau}} \equiv \sqrt{(p_{j_\tau}+p_{\tilde{\tau}})^2}$,
where $p_{j_\tau}$ and $p_{\tilde{\tau}}$ are four-momenta of $\tau$-jet
and $\tilde{\tau}$, respectively, there should be an upper endpoint at
$m_{\chi^0_i}$.

Once the lightest neutralino mass is measured, it can be used to
reconstruct the slepton masses $m_{\tilde{e}_R}$ and
$m_{\tilde{\mu}_R}$.  For this purpose, we use the decay process
$\chi^0_1\rightarrow l^\pm\tilde{l}_R^\mp$ (with $l=e$ or $\mu$)
followed by $\tilde{l}_R^\mp\rightarrow l^\mp\tau\tilde{\tau}_1$ and
$\tau\rightarrow j_\tau\nu_\tau$.  
Since the $\tau$ leptons from the $\tilde{l}^\mp_R$ decay are expected
to be highly boosted, the directions of $\tau$ and $\tau$-jet are
(almost) aligned.
Then, the four momentum of $\tau$ is obtained by requiring
$(p_{l^+}+p_{l^-}+p_{\tau}+p_{\tilde{\tau}})^2=m_{\chi^0_1}^2$.  Once
$p_{\tau}$ is known, we define
$M_{\tilde{l}^\pm}\equiv\sqrt{(p_l^\pm+p_{\tau}+p_{\tilde{\tau}})^2}$.
Because one of the charged leptons is from the decay of
$\tilde{l}_R^\pm$, $M_{\tilde{l}^+}$ or $M_{\tilde{l}^-}$ is equal to
$m_{\tilde{l}_R}$.  
Therefore, we expect to have a sharp peak at the slepton masses in the
distribution of $M_{\tilde{l}^\pm}$.

In order to reconstruct the neutralino and slepton masses with the
above-mentioned procedure, we need enough amount of neutralinos to
perform the statistical analysis. 
The production cross section of neutralinos, of course, depends on the
model parameters, but in many cases the neutralinos are copiously
produced from the decay of squarks.
In fact, squark mass measurements may also be possible by reconstructing
the decay products as we demonstrate later.

In Ref.~\cite{Hinchliffe:1998ys}, the squark mass measurement was
discussed for the case that $\tilde{e}_R$ and $\tilde{\mu}_R$, as well
as $\tilde{\tau}_1$, are long-lived.  In such a case, the momentum of
$\tilde{l}_R$ can be well measured.  By using the decay chain
$\tilde{q}\rightarrow q\chi^0_1\rightarrow q l\tilde{l}_R$, it was
pointed out that a selection based on the invariant mass of the system
$(l^\pm, \tilde{l}_R^\mp)$ works to find $l$ and $\tilde{l}_R$ which
are from the same neutralino.  Then, the invariant mass of the system
$(j, l, \tilde{l}_R)$ (with $j$ being one of high-$p_T$ jet) directly
provides the squark mass.

If $\tilde{l}_R$ decays in the detector, this method is not
applicable.  Even in such a case, however, the squark mass information
can be obtained by reconstructing the momentum of $\tau$.  Considering
the lightest neutralino production process $\tilde{q}\rightarrow
q\chi^0_1\rightarrow q\tau\tilde{\tau}_1$ followed by hadronic decay
of $\tau$, and using the fact that $m_{\chi^0_1}$ can be determined
from the endpoint analysis, we can first reconstruct the momentum of
$\tau$ by requiring that the invariant mass
$\sqrt{(p_{\tau}+p_{\tilde{\tau}})^2}$ be equal to $m_{\chi^0_1}$.
Then, the squark mass can be obtained by
$\sqrt{(p_q+p_{\tau}+p_{\tilde{\tau}})^2}$, where $p_q$ is the
momentum of the quark from the squark decay, whose information is in
principle imprinted in the momentum of one of the observed jets.  We
note here that the flavor information on the primary quark is hardly
obtained except for the $b$-jet, so we can perform only the
flavor-blind analysis.  In a large class of models, however, first-
and second-generation squarks are almost degenerate.  (For example, in
the GMSB and mSUGRA models, this is the case.)  In such a case, as we
will see in the next section, a clear peak corresponding to the squark
mass is obtained even though we cannot specify the flavor of the jets.

Of course, in the actual situation, the measurements of the masses are
not straightforward. This is because (i) the momenta of the decay
products of $\chi^0_1$ are measured with some uncertainties, (ii) jets
with small multiplicity (i.e., $\tau$-jet-like objects) are also
produced by the QCD process, which mimics the hadronically decaying
$\tau$-lepton, and (iii) combinatorial backgrounds should exist.  In
the next section, we discuss how well our idea works at the LHC
experiment using MC analysis, taking account of the effects of (i) $-$
(iii).

\section{Numerical Analysis}
\label{sec:mc}
\setcounter{equation}{0}

We demonstrate in the following the method to measure the sparticle
masses presented in the previous section by performing a Monte Carlo
simulation. As a model with the spectrum in Eq.~\eqref{MassRelation},
we use the minimal GMSB model.  The superparticle spectrum is
parametrized by $\Lambda$ (the ratio of the $F$- and $A$-components of
the SUSY breaking field), $M_{\rm mess}$ (the messenger scale),
$N_{\bf 5}$ (the number of messenger multiplet in units of ${\bf
  5}+{\bf \bar{5}}$ representation), $\tan\beta$ (the ratio of the
vacuum expectation values of two Higgs bosons), and the sign of $\mu$
(the SUSY invariant Higgs mass).
We take
\begin{eqnarray}
  \Lambda = 60\ {\rm TeV}, \ \ 
  M_{\rm mess} = 900\ {\rm TeV}, \ \ 
  N_{\bf 5} = 3,\ \ 
  \tan\beta = 35,\ \ 
  {\rm sign} (\mu) = +.
\end{eqnarray}
Even though we adopt specific models for our analysis, it should be
noted that our procedure works in a wider class of models as far as
the mass relation \eqref{MassRelation} holds.

The mass spectrum of superparticles is calculated by using ISAJET
7.64~\cite{Paige:2003mg}.  In order to see how effective our method
is, we reduce the mass of $\tilde{\mu}_R$ by $1\ {\rm GeV}$.  With the
present choices of parameters, the MSSM-LSP is $\tilde{\tau}_1$, and
the lightest neutralino, which is almost the Bino, is heavier than
lighter sleptons $\tilde{e}_R$ and $\tilde{\mu}_R$.  The mass spectrum
of the superparticle is summarized in Table~\ref{table:susymass}.  The
branching ratios of the decay of $\chi^0_1$ are given by
$Br(\chi^0_1\rightarrow \tau\tilde{\tau}_1)= 61.4$ \% and
$Br(\chi^0_1\rightarrow l\tilde{l}_R) = 19.3$ \%, while $\tilde{l}_R$
dominantly decays as $\tilde{l}_R\rightarrow l\tau\tilde{\tau}_1$
($\simeq$ 100 \%).

\begin{table}
  \centering
  \begin{tabular}{cr} 
    \hline \hline
    Particle & Mass (GeV) \\
    \hline
    $\tilde{g}$      & 1309.39  \\
    $\tilde{u}_{L}$  & 1231.70   \\
    $\tilde{u}_{R}$  & 1183.97   \\
    $\tilde{d}_{L}$  & 1234.28   \\
    $\tilde{d}_{R}$  & 1180.19   \\
    $\tilde{t}_{1}$  & 1082.85 \\
    $\tilde{t}_{2}$  & 1195.08 \\
    $\tilde{b}_{1}$  & 1145.24 \\
    $\tilde{b}_{2}$  & 1185.83 \\
    $\tilde{\nu_l}$  & 388.05 \\
    $\tilde{l}_{L}$  & 396.19 \\
    $\tilde{\tau}_{2}$ & 402.57 \\
    $\tilde{\nu}_\tau$ &  383.80 \\
    $\tilde{e}_{R}$  & 194.39 \\
    $\tilde{\mu}_{R}$  & 193.39 \\
    $\tilde{\tau}_{1}$ & 148.83 \\
    $\chi^0_{1}$ & 239.52 \\
    $\chi^0_{2}$ & 425.92 \\
    $\chi^0_{3}$ & 508.41 \\
    $\chi^0_{4}$ & 548.67 \\
    $\chi^\pm_{1}$ & 425.45 \\ 
    $\chi^\pm_{2}$ & 548.43 \\
    $h$ & 115.01 \\ 
    \hline\hline
  \end{tabular}
  \caption{Masses of the superparticles and the lightest 
    Higgs boson $h$ in units of GeV.  The input  parameters are
    $\Lambda = 60\ {\rm TeV}$, $M_{\rm mess} = 900\ {\rm TeV}$, $N_{\bf 5} = 3$, 
    $\tan\beta = 35$, ${\rm sign} (\mu) = +$.
    (We use the top-quark mass of $171.3\ {\rm GeV}$.)
    We reduced the mass of $\tilde{\mu}_R$ by $1\ {\rm GeV}$.}
  \label{table:susymass}
\end{table}

We have generated 66960 SUSY events corresponding to the luminosity of
100~${\rm fb}^{-1}$ at a $pp$ collider with the center of mass energy
of $14\ {\rm TeV}$ by using the HERWIG 6.510
package~\cite{Corcella:2000bw,Moretti:2002eu}. (The total cross section
of the SUSY events is $669.6\ {\rm fb}$.)
Events are passed through the PGS4 detector simulator~\cite{PGS4}.\footnote
{In PGS, effects of energy leakage into the hadronic calorimeter are
  included for electromagnetic objects (i.e., $e^\pm$ and $\gamma$),
  which results in an underestimation of the energy of $e^\pm$.  We
  expect that the energy of $e^\pm$ will be calibrated by requiring
  that the $Z$-boson mass is well reconstructed.  So, we estimate the
  observed energy of $e^\pm$ by neglecting the energy leakage.}
The momentum resolution of $\tilde{\tau}_1$ is assumed to be the same
as those of muons as long as $0.4\leq\beta_{\tilde{\tau}_1}\leq 0.9$.

\subsection{Neutralino masses}

\begin{figure}[t]
 \centerline{\epsfxsize=\textwidth\epsfbox{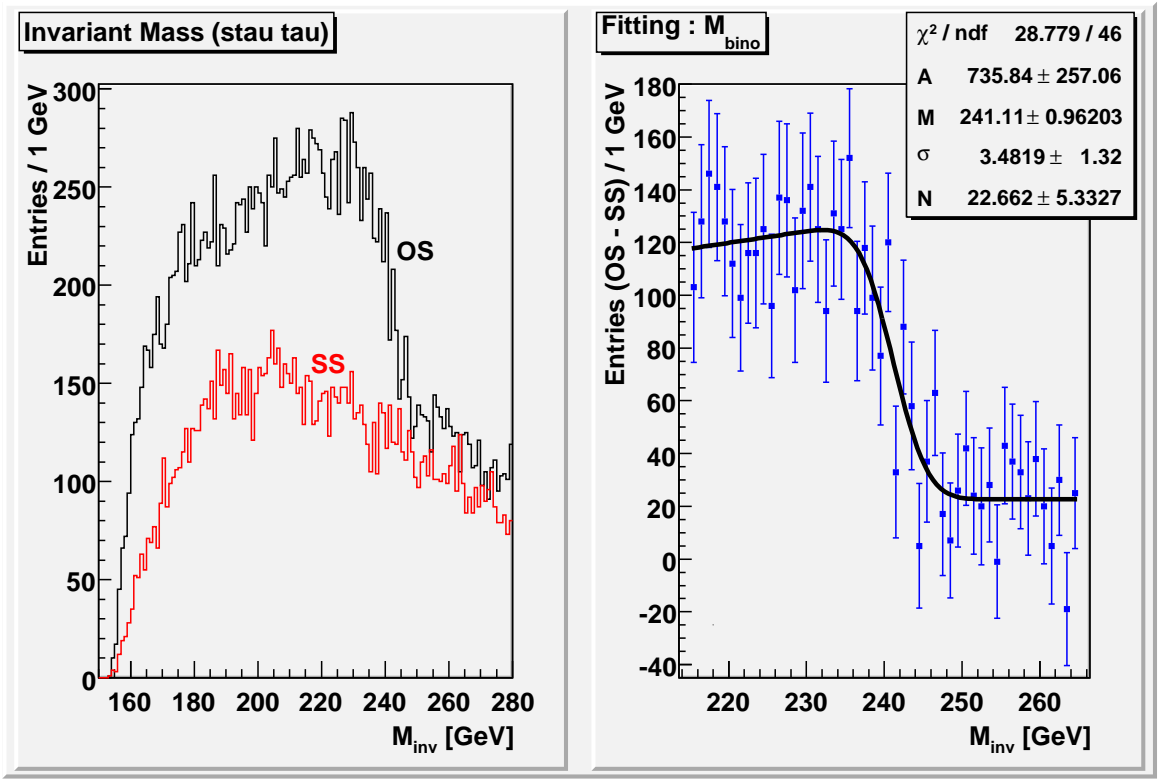}}
 \caption{\small [Left]: The invariant mass distributions of opposite
   sign (black) and same sign (red) pairs for the GMSB case.  Here, we
   take ${\cal L}=100\ {\rm fb}^{-1}$, and the postulated stau mass is
   taken to be the underlying value. [Right]: The fitting of the
   lightest neutralino mass after the charge subtraction.}
 \label{fig:Mbino_gmsb}
\end{figure}

We first discuss the neutralino mass measurement explained in the
previous section. We identify a decay process:
\begin{eqnarray}
  \chi^0_1\rightarrow \tau^\pm \tilde{\tau}_1^\mp,
\end{eqnarray}
followed by the hadronic decay of the $\tau$-lepton.\footnote
{We have also considered the possibility to use the leptonic decay
  mode of $\tau$-lepton.  However, for the signals from the leptonic
  decay mode, the combinatorial background is so severe that the
  results are much worse than the case with the hadronic decay mode of
  $\tau$.}
The following selection cuts are applied:
\begin{itemize}
\item[1a)] At least one $\tilde{\tau}_1$ with the velocity
  $0.4\leq\beta_{\tilde{\tau}_1}\leq 0.9$; such $\tilde{\tau}_1$ is
  assumed to be detected with the efficiency of 100\ \% with no
  standard-model background.  In the study of the invariant-mass
  distribution, $\tilde{\tau}_1$ with the velocity in this range are
  used.
\item[1b)] At least one $\tau$-tagged jet with $p_T>15\ {\rm
    GeV}$, which is denoted as $j_\tau$.
\end{itemize}
By using events passed the selection cuts, we calculate the invariant
mass for all the possible pairs of $(j_\tau,\tilde{\tau}_1)$:
\begin{eqnarray}
  M_{j_\tau\tilde{\tau}_1} \equiv \sqrt{(p_{j_\tau}+p_{\tilde{\tau}})^2}.
\end{eqnarray}
The charges of $j_\tau$ and $\tilde{\tau}_1$, which are both observable,
should be opposite for signal events; we call such events as
opposite-sign (OS) event.
In Fig.\ \ref{fig:Mbino_gmsb} (left), we plot the distribution of the
invariant mass $M_{j_\tau\tilde{\tau}_1}$.  One can find a sharp
drop-off at $M_{\tilde{\tau}_1j_\tau}\sim 240\ {\rm GeV}$, which is
close to the input value of the lightest neutralino mass.  However,
one can also see a long tail above the drop-off.  Those backgrounds
come from fake $\tau$-jets (mis-identified QCD jets) as well as from
wrong combination where $\tau$ and $\tilde{\tau}$ have different
parents.  Since those backgrounds are charge-blind, their
contributions to the OS and same-sign (SS) histograms are expected to
be the same amount.
In Fig.\ \ref{fig:Mbino_gmsb} (left), we also show the distribution of
the invariant mass for the SS event.  One can see that the number of
OS events is significantly larger than that of SS events for
$M_{\tilde{\tau}_1j_\tau}\lesssim 240\ {\rm GeV}$, and those two
become comparable for a larger invariant mass.  This fact confirms our
expectation that the tail is mostly from the fake $\tau$-jets and
wrong combination.

By taking a difference between the OS and SS events, one can subtract
the contributions from the backgrounds.
The distribution of the signal events after the charge subtraction is
shown in Fig.\ \ref{fig:Mbino_gmsb} (right).  We can see a clearer
endpoint.
We search the upper endpoint by fitting the histogram with the
Gaussian-smeared triangular function (including the effect of
background)~\cite{Bachacou:1999zb}:
\begin{eqnarray}
  \mbox{(Number of events)} = 
  A \int_{-1}^1 dz \exp 
  \left[ \frac{-1}{2\sigma^2}
    \left( 
      M_{\tilde{\tau}_1j_\tau}
      - M^{\rm (max)} \sqrt{\frac{1+z}{2}}
    \right)^2
  \right]
  + N_{\rm BG},
\end{eqnarray}
where $A$, $M^{\rm (max)}$, $\sigma$, and $N_{\rm BG}$ are fitting
parameters; in particular, $M^{\rm (max)}$ corresponds to the upper
endpoint. The endpoint is determined to be $(241.11\pm 0.96)\ {\rm GeV}$,
which is consistent with the underlying value of $m_{\chi^0_1}$.

\begin{figure}[t]
  \centerline{\epsfxsize=\textwidth\epsfbox{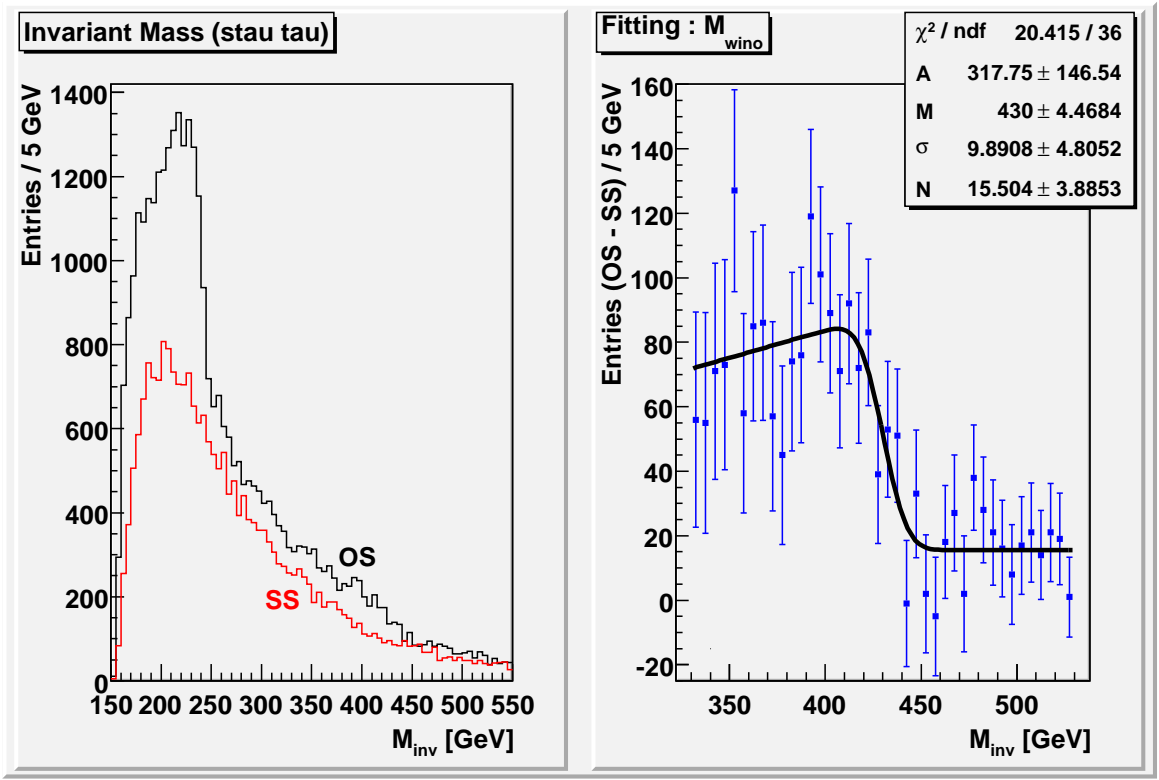}}
  \caption{\small [Left]: The invariant mass distributions of opposite
    sign (black) and same sign (red) pairs for the GMSB case.  Here,
    we take ${\cal L}=100\ {\rm fb}^{-1}$, and the width of the bin is
    $10\ {\rm GeV}$. [Right]: The fitting of the second-lightest
    neutralino mass after charge subtraction.}
  \label{fig:Mwino_gmsb}
\end{figure}

In the above study, the charge subtraction method was very powerful in
finding the neutralino endpoint.
One can further try to find the second-lightest neutralino,
which is almost the neutral Wino in this model.
Although no clear endpoint can be seen in the OS events at $\sim
m_{\chi^0_2}$, it shows up after the subtraction.  (See Fig.\
\ref{fig:Mwino_gmsb}, where the distribution of
$M_{j_\tau\tilde{\tau}_1}$ is shown with the width of the bin of $5\
{\rm GeV}$.)
By fitting the edge with the same function, $m_{\chi^0_2}$ is measured
to be $(430.0\pm 4.5)\ {\rm GeV}$, which is slightly larger than
the underlying value.

Before closing this subsection, we comment on the
$m_{\tilde{\tau}_1}$-dependence of the reconstructed mass.  The stau
mass is expected to be measured in the long-lived stau scenario by
combining the velocity and the momentum informations on the
$\tilde{\tau}_1$ track; the expected error in the stau mass measurement
is $\delta m_{\tilde{\tau}_1}\sim O(100\ {\rm
MeV})$~\cite{Ambrosanio:2000ik, ellis_atlnote}.
The neutralino mass measurements will therefore be affected by
$O(100~{\rm MeV})$, which is less significant compared to the statistical
uncertainties.

\subsection{Slepton masses}

We proceed to the discussion of the slepton mass measurements.  In
particular, it is interesting to measure the masses of $\tilde{e}_R$
and $\tilde{\mu}_R$ since one can learn the flavor structure of the
model.
Note that study of those particles are difficult in the Bino LSP
scenario since they do not appear in the cascade decays.
In order to measure $m_{\tilde{l}_R}$, we use the decay chain
\begin{eqnarray}
  \chi^0_1 \rightarrow l^\pm \tilde{l}_R^\mp
  \rightarrow l^\pm l^\mp \tau \tilde{\tau}_1,
  \label{decay12}
\end{eqnarray}
followed by the hadronic decay of $\tau$.  In the decay chain
\eqref{decay12}, the charges of $\tau$ and $\tilde{\tau}$ are
opposite.  We apply the following selections:
\begin{itemize}
\item[2a)] At least one $\tilde{\tau}_1$ with the velocity
  $0.4\leq\beta_{\tilde{\tau}_1}\leq 0.9$.
\item[2b)] At least one $\tau$-tagged jet with $p_T>15\ {\rm
    GeV}$.
\item[2c)] At least one pair of isolated opposite-charge same-flavor
  leptons.  We require $p_T>15\ {\rm GeV}$ for leptons.
\end{itemize}
For each event, we choose all the possible combinations of $(l^+, l^-,
j_\tau, \tilde{\tau}_1)$ with requiring charges of $\tilde{\tau}_1$ and
$j_\tau$ to be opposite.  Assuming that $l^+$, $l^-$, $j_\tau$, and
$\tilde{\tau}_1$ are all from the lightest neutralino whose mass is
determined already, we can calculate the four-momentum of the
$\tau$-lepton.  Because the $\tau$-lepton from the slepton decay is
usually ultra-relativistic, $\tau$ and $j_\tau$ are expected to be
emitted to (almost) the same direction.  The four-momentum of $\tau$ can
thus be estimated as
\begin{eqnarray}
  p_\tau = z^{-1} p_{j_\tau},
\end{eqnarray}
where
\begin{eqnarray}
  z = 
  \frac{2 p_{j_\tau} \cdot (p_{l^+} + p_{l^-}+ p_{\tilde{\tau}})}
  {\tilde{m}_{\chi^0_1}^2 - (p_{l^+} + p_{l^-} + p_{\tilde{\tau}})^2},
  \label{zdef}
\end{eqnarray}
with $\tilde{m}_{\chi^0_1}$ being the postulated value of the lightest
neutralino mass for the analysis.  The events with $z>1$ are rejected.
In order to reconstruct the slepton mass $m_{\tilde{l}_R}$, we study
the distribution of the following quantity:
\begin{eqnarray}
  M_{\tilde{l}} = 
  \sqrt{(p_{l}+p_{\tilde{\tau}}+z^{-1}p_{j_\tau})^2}.
\end{eqnarray}
Here, $p_{l}$ is the four momentum of one of two charged leptons in
$(l^+, l^-, j_\tau, \tilde{\tau}_1)$.  Because we cannot tell
which lepton is the one from the slepton decay, we calculate the
invariant masses by using both of the possibilities.

\begin{figure}[t]
  \centerline{\epsfxsize=0.85\textwidth\epsfbox{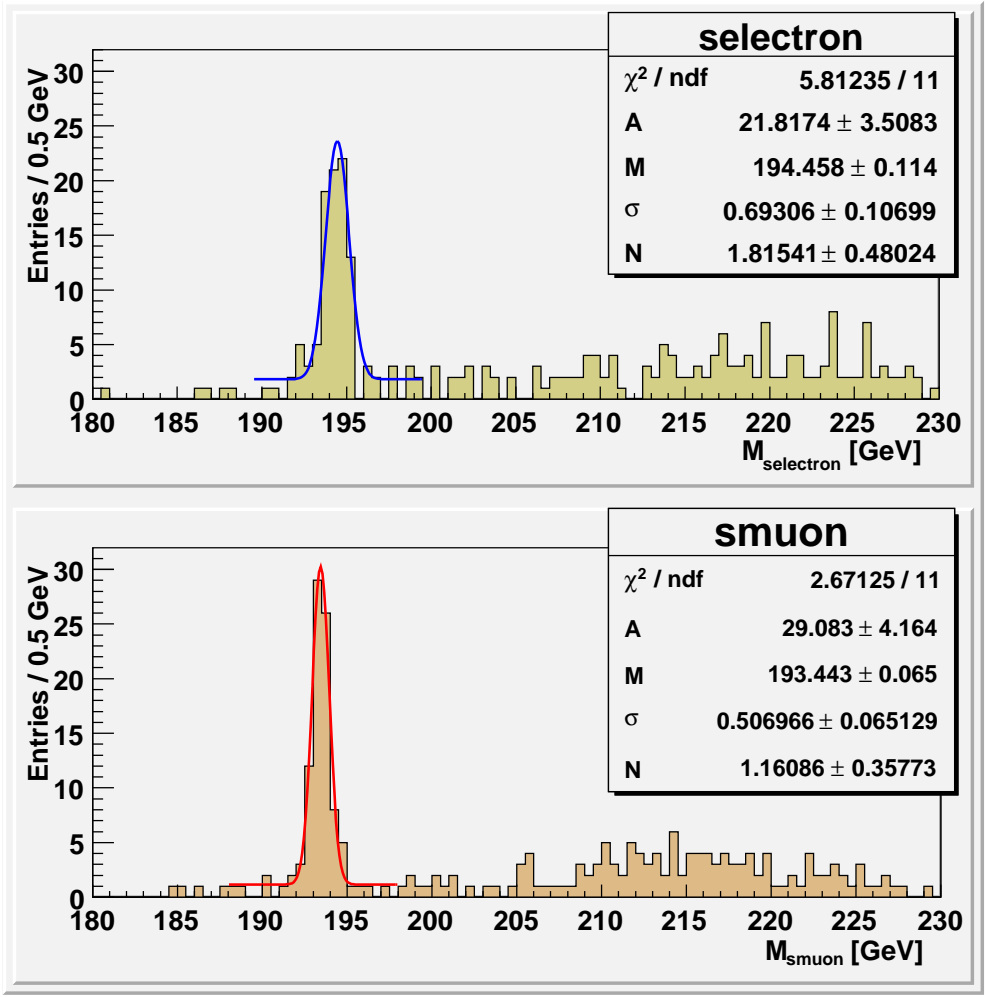}}
  \caption{Distribution of $M_{\tilde{l}}$ with $l=e$ (left) and $\mu$
    (right) for the GMSB case with ${\cal L}=100\ {\rm
      fb}^{-1}$. Here, the postulated stau and neutralino masses are
    chosen to be the underlying values.}
\label{fig:Msl_gmsb}
\end{figure}

In Fig.\ \ref{fig:Msl_gmsb}, we show the distributions of
$M_{\tilde{l}}$ with $l=e$ and $\mu$.  Here, the postulated values of
the stau and the lightest neutralino masses are taken to be equal to
the underlying values.
Even though the distribution contains the combinatorial background,
one can see sharp peaks at the slepton masses.
Fitting the peaks with the Gaussian function plus a constant
background,
\begin{eqnarray}
  \mbox{(Number of events)} = 
  A \exp 
  \left[ - \frac{1}{2\sigma^2}
    \left( 
      M_{\tilde{l}} 
      - M^{\rm (peak)} 
    \right)^2
  \right]
  + N_{\rm BG},
\end{eqnarray}
with $A$, $M^{\rm (peak)}$, $\sigma$, and $N_{\rm BG}$ being parameters,
the peak positions $M^{\rm (peak)}$ are determined to be $(194.46\pm
0.11)~{\rm GeV}$ and $(193.44\pm 0.07)~{\rm GeV}$ for $l=e$ and $\mu$,
respectively.  Those values are in very good agreements with the actual
slepton masses.

\begin{figure}[t]
  \centerline{\epsfxsize=\textwidth\epsfbox{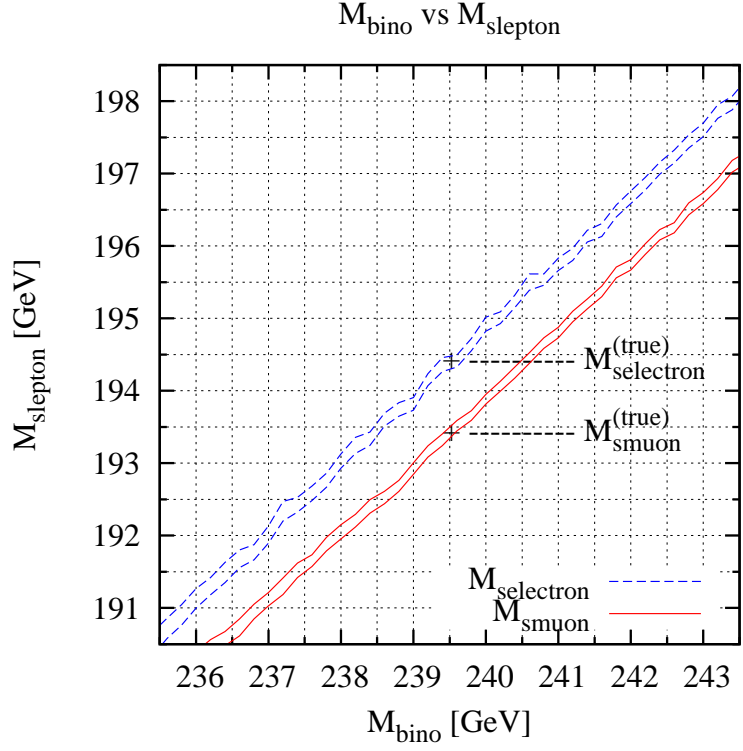}}
  \caption{The reconstructed slepton masses as functions of the
    postulated lightest neutralino mass for the GMSB case.}
\label{fig:MbinoMslep_gmsb}
\end{figure}

In the slepton mass determination, we should consider errors
associated with the uncertainties in the mass measurements of stau and
the lightest neutralino.  In particular, the reconstructed slepton
masses are sensitive to the uncertainty in the lightest neutralino
mass.  In Fig.\ \ref{fig:MbinoMslep_gmsb}, we show the reconstructed
slepton masses (i.e., the 1-$\sigma$ upper and lower bounds with
${\cal L}=100\ {\rm fb}^{-1}$) as functions of the postulated lightest
neutralino mass.  One can see that the reconstructed slepton masses
depend linearly on the postulated lightest neutralino mass.
As mentioned in the previous section, the lightest neutralino mass can
be measured by the endpoint study, and the uncertainty is $\sim 1 \
{\rm GeV}$, which gives the error of $[ \delta
m_{\tilde{l}_R}]_{\delta m_{\chi^0_1}}\sim 1\ {\rm GeV}$ in the
slepton mass measurement.  Importantly, however, the mass difference
$m_{\tilde{e}_R}-m_{\tilde{\mu}_R}$ is insensitive to the neutralino
mass.
The dependence of the reconstructed slepton masses on the postulated
stau mass is proportional to
$(m_{\chi^0_1}-m_{\tilde{l}_R})/m_{\chi^0_1}$, and thus rather weak in
the present model.

\subsection{Squark mass}

Sqark masses can be also measured by reconstructing the decay chain:
\begin{eqnarray}
  \tilde{q} \rightarrow q \chi^0_1 \rightarrow q \tau \tilde{\tau},
  \label{decaysquark}
\end{eqnarray}
followed by hadronic decay of $\tau$.  In order to use the decay chain
\eqref{decaysquark}, we adopt the following event selections:
\begin{itemize}
\item[3a)] At least one $\tilde{\tau}_1$ with the velocity
  $0.4\leq\beta_{\tilde{\tau}_1}\leq 0.9$.
\item[3b)] At least one jet with $p_T> 100\ {\rm GeV}$.
\item[3c)] At least one $\tau$-tagged jet with $p_T>15\ {\rm
    GeV}$.
\item[3d)] No isolated lepton with $p_T>15\ {\rm GeV}$.
\end{itemize}
The condition 3d) is to eliminate the mis-reconstruction of
leptonically decaying $\tau$.

Then, we consider all the possible combinations $(j, j_\tau,
\tilde{\tau}_1)$, where $j$ is one of four leading jets with $p_T> 100\
{\rm GeV}$.  Assuming that $j_\tau$ and $\tilde{\tau}_1$ are both from
the decay of the neutralino, we first reconstruct the momentum of
$\tau$, which is given by
\begin{eqnarray}
  p_\tau = z_{\tilde{q}}^{-1} p_{j_\tau},
\end{eqnarray}
where
\begin{eqnarray}
  z_{\tilde{q}} = 
  \frac{2p_{j_\tau} \cdot p_{\tilde{\tau}}}
  {\tilde{m}_{\chi^0_1}^2 - m_{\tilde{\tau}}^2}.
  \label{zqdef}
\end{eqnarray}
Combinations with $z_{\tilde{q}}>1$ is eliminated.  Then, we study the
distribution of the following variable:
\begin{eqnarray}
  M_{\tilde{q}} = 
  \sqrt{(p_{j}+p_{\tilde{\tau}}+z_{\tilde{q}}^{-1}p_{j_\tau})^2}.
\end{eqnarray}

\begin{figure}[t]
  \centerline{\epsfxsize=\textwidth\epsfbox{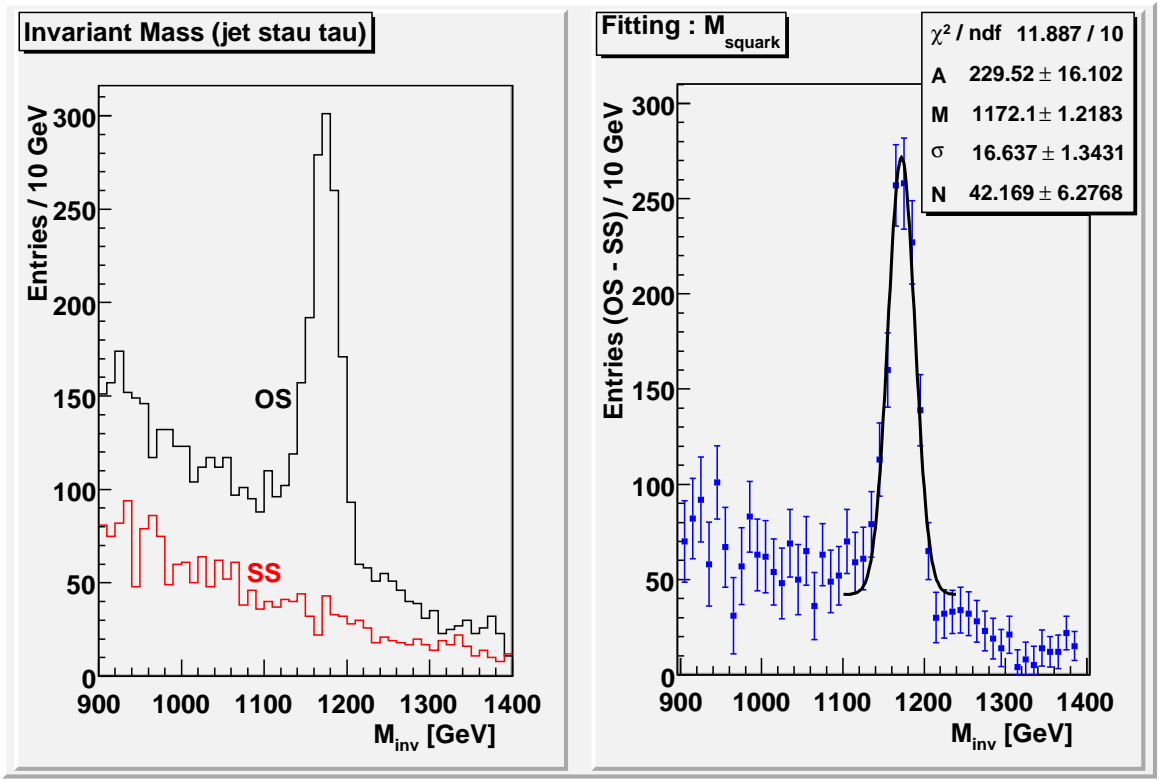}}
  \caption{\small [Left]: The invariant mass distributions of opposite
    sign (black) and same sign (red) events.  Here, we take ${\cal
      L}=100\ {\rm fb}^{-1}$, and the postulated stau mass is taken to
    be the underlying value of the stau mass. [Right]: The fit of the
    number of events with the Gaussian function.}
  \label{fig:Msquark_gmsb}
\end{figure}

The distribution of $M_{\tilde{q}}$ is shown in Figs.\
\ref{fig:Msquark_gmsb} for OS and SS events.  We can see a clear peak
for the OS event at the position corresponding to the squark masses,
while no clear peak is seen for the SS event.  By fitting the
histogram after the charge subtraction with the Gaussian function, the
peak position is obtained to be $m_{\tilde{q}}=(1172.1\ \pm 1.2)\ {\rm
  GeV}$, while underlying squark masses are $m_{\tilde{u}_R}=1184\
{\rm GeV}$, $m_{\tilde{d}_R}=1180\ {\rm GeV}$, $m_{\tilde{u}_L}=1232\
{\rm GeV}$, and $m_{\tilde{d}_L}=1234\ {\rm GeV}$.  The dominant
contribution to the histogram is from right-handed squarks. Compared
to the right-handed squark masses, the observed peak location is
smaller than the underlying values by $\sim 10\ {\rm GeV}$.  It is
partly from the mis-measurement of the jet energy and also from
effects of the background.
The uncertainties from the mis-measurements of the Bino and the stau
masses are much smaller than the statistical error as far as the
luminosity of ${\cal L}\sim 100\ {\rm fb}^{-1}$ is used.

In principle, a similar analysis can be performed by using the
second-lightest neutralino (which is almost the Wino), which should
provide information on the left-handed squark mass.  However, the
left-handed squarks decay into chargino (plus jet) as well as into
neutralino.  In addition, it is difficult to distinguish the left- and
right-handed squark production events in the event-by-event basis.
These become the source of extra background and reduce the
signal-to-background ratio.  Thus, for the study of the left-handed
squark mass, we should perform more sophisticated analysis.  A
detailed discussion will be given elsewhere \cite{IKM_InPreparation}.

\section{Implications}
\label{sec:imp}

We have seen that, in the long-lived stau scenario, masses of various
SUSY particles are expected to be measured with a very good accuracy
at the LHC.
This level of precise deterimnation of the mass spectrum may reveal
the origin of SUSY breaking terms.

As we have seen, the masses of the lightest and the
second-lightest neutralino, which almost correspond to the Bino
and the neutral Wino, respectively, can be determined with the
endpoint analyses.  Then, combined with the gluino-mass information
from, for example, the cross-section information for the process
$pp\rightarrow\tilde{g}\tilde{g}$, we can test the GUT relation among
the gaugino masses.

The masses of $\tilde{e}_R$ and $\tilde{\mu}_R$ are also measured with
good accuracies.  Even though the mass determinations of $\tilde{e}_R$
and $\tilde{\mu}_R$ are sensitive to the uncertainty in the lightest
neutralino mass, the measured value of the mass differene
$m_{\tilde{e}_R}-m_{\tilde{\mu}_R}$ is not affected to it.  The mass
difference $m_{\tilde{e}_R}-m_{\tilde{\mu}_R}$ contains various
informations.  Neglecting the flavor mixing, the slepton mass matrix has
the form:
\begin{eqnarray}
  {\cal M}^2_{\tilde{l}} = 
  \left(
    \begin{array}{cc}
      m_L^2 + \Delta_{\tilde{l}, {LL}} & \Delta_{\tilde{l}, {LR}} \\
      \Delta_{\tilde{l}, {LR}} & m_R^2 + \Delta_{\tilde{l}, {RR}}
    \end{array} \right),
\end{eqnarray}
where $m_L$ and $m_R$ are independent of the flavor index.  Then,
assuming that $\Delta\ll m_{L,R}^2$, the mass difference is given by
\begin{eqnarray}
  m_{\tilde{e}_R} - m_{\tilde{\mu}_R} \simeq
  \frac{1}{2m_R} (\Delta_{\tilde{e}, {RR}} - \Delta_{\tilde{\mu}, {RR}})
  - \frac{1}{2m_R (m_L^2 - m_R^2)} 
  (\Delta_{\tilde{e}, {LR}}^2 - \Delta_{\tilde{\mu}, {LR}}^2).
\end{eqnarray}
Thus, the slepton-mass difference is sensitive to the left-right
mixing, whose information is in $\Delta_{\tilde{l}, {LR}}$, as well as
to the difference of the diagonal elements of the mass matrix.  Even
if the dominant contribution to the masses of sleptons (with the same
gauge quantum numbers) is almost universal at some high energy scale,
the universality is affected by various effects.  

For example, in the GMSB models, the non-universality is originated
from the supergravity effects and also from one-loop corrections with
the muon Yukawa interaction.  The supergravity effects are estimated
as $\Delta_{\tilde{l},LL}\sim\Delta_{\tilde{l},RR}\sim O(m_{3/2}^2)$,
with $m_{3/2}$ being the gravitino mass.  Thus, if the gravitino mass
is as large as a few GeV, which is merginally consistent with the
flavor-violation constraints depending on other SUSY parameters, the
supergravity effect may be seen assuming that the mass difference
$m_{\tilde{e}_R}-m_{\tilde{\mu}_R}$ is determined with the accuracy of
$O(100\ {\rm MeV})$.

\begin{figure}[t]
\begin{center}
\includegraphics[width=10cm]{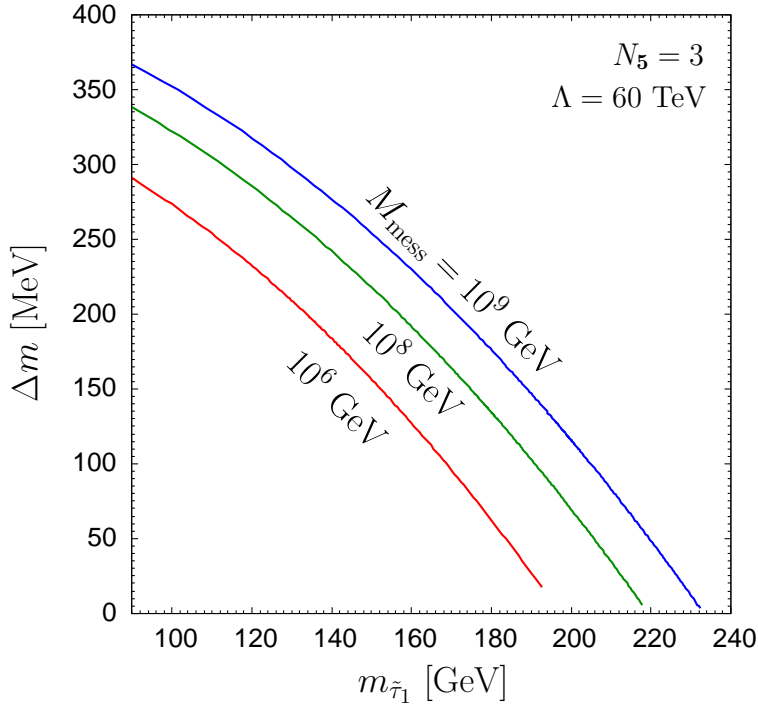} 
\end{center}
\caption{\small The selectron-smuon mass splitting calculated in the
 sweet-spot model.}
\label{fig:sweet}
\end{figure}

The size of the loop effects is enhanced when $\tan\beta$ is large,
and the contribution to the mass splitting is roughly estimated to be
$O(100\ {\rm MeV})$ for $\tan\beta\gtrsim 30$.  The actual value of
the loop effects depends on the mechanism to generate the Higgs mass
parameters since there is a one-loop diagram with Higgs fields in the
loop.  As an example of explicit models we calculate the mass
splitting by solving renormalization group (RG) equations in the
sweet-spot supersymmetry model~\cite{Ibe:2007km}.  SUSY breaking terms
are determined by the parameters: $\Lambda$, $M_{\rm mess}$, $N_{\bf
  5}$, and $\mu$. The Higgs mass parameters are generated at the
unification scale rather than the messenger scale in this model, and
hence there is a large logarithm in the quantum corrections. Also, the
$\tan \beta$ parameter is predicted to be large due to the boundary
condition at the messenger scale, $B=0$. The Yukawa coupling constants
for the charged leptons are therfore enhanced. For these reasons, the
mass splitting, $\Delta m \equiv m_{\tilde e_R} - m_{\tilde \mu_R}$,
is expected to be large. Note also that for the same reason, $\tilde
\tau_1$ can be significantly lighter than the other sleptons, that
enables us to perform the analysis in the previous section.

The mass spectrum is calculated with fixing $\Lambda$ and $N_{\bf 5}$
as
\begin{eqnarray}
\Lambda = 60\ {\rm TeV}, \ \ 
N_{\bf 5}=3,
\end{eqnarray}
so that the gaugino masses are similar to the ones in the example we
took in the previous section. Since the value of $\mu$ (the Higgsino
mass parameter) is strongly correlated to the stau mass in this model
(through the RG effects mentioned above), we can trade the input
parameter $\mu$ with $m_{\tilde \tau_1}$.
We show in Fig.~\ref{fig:sweet} the mass splitting as a function of the
stau mass for various values of the messenger scale.
For $M_{\rm mess} \gtrsim 10^{10}$~GeV, the slepton is heavier than the
Bino with this set of parameters.
The lines are terminated at the point where the stau becomes heavier
than the other sleptons.
One can see that the mass splitting can be as large as $\sim 300~{\rm
  MeV}$ which is large enough to be observed at the LHC.
This calculation demonstrates that the mass splitting has a rich
information on the microscopic theory to generate the SUSY breaking
terms, especially those for the Higgs fields.
It is also interesting to note that the LHC may be able to see an effect
of the muon Yukawa coupling.

\section{Conclusions}
\label{sec:conclusions}
\setcounter{equation}{0}

We have developed methods to measure the superparticle masses at the LHC
in the long-lived stau scenario.  
We have concentrated on the scenario where
$m_{\tilde{\tau}_1}<m_{\tilde{l}_R}<m_{\chi^0_1}$ (with $l=e$ and $\mu$)
and demonstrated that the masses of neutralinos, sleptons and squarks
can be well determined by endpoint or peak analysis.

In the neutralino mass measurements by the endpoint analysis of the
$M_{j_\tau\tilde{\tau}}$ invariant masses, we have seen that the charge
  subtraction method is useful to identify the endpoints.
In the sample point we have adopted, the estimated
error in the lightest neutralino mass measurement is $\sim 1\ {\rm
GeV}$.  We have shown that even the second-lightest
neutralino mass can be measured with an accuracy of $ 4-5 \ {\rm GeV}$.

Once the lightest neutralino mass is known, it can be used to
determine the masses of selectron and smuon.  With the decay chain
$\chi^0_1\rightarrow l^\pm\tilde{l}_R^\mp\rightarrow l^\pm
l^\mp\tau\tilde{\tau}_1$, followed by hadronic decay of $\tau$, we
first reconstruct the momentum of $\tau$ by using the relation
$m_{\chi^0_1}=\sqrt{(p_{l^+}+p_{l^-}+p_{\tau}+p_{\tilde{\tau}})^2}$.
Then, the slepton mass $m_{\tilde{l}_R}$ is determined in each event
up to a combinatorics.  With this method, we have seen that very sharp
peaks are obtained around the underlying values of the slepton masses,
which gives precise determination of the slepton masses.  We have
estimated the error in the slepton mass determination, which is $\sim
100\ {\rm MeV}$.  Such a precise measurement of the slepton mass
enables us to study effects of renormalization group, supergravity,
and/or left-right mixing on the slepton masses.  We also demonstrated
that a sharp peak corresponding to (right-handed) squarks can be
observed by using the $\tilde q \to q \chi_1^0 \to q \tau\tilde
\tau_1$ events.

\noindent
{\it Acknowledgments:}
We thank Prof.\ S. Asai for useful discussion.  This work was
supported in part by the Grant-in-Aid for Scientific Research from the
Ministry of Education, Science, Sports, and Culture of Japan, no.\
21840006 (R.K.)  and no.\ 19540255 (T.M.).

\end{document}